\documentstyle[12pt,epsf]{article}
\begin{document}
\title{Systematic and statistical errors in homodyne measurements 
of the density matrix}
\author{G. M. D'Ariano$^{a}$, C. Macchiavello$^{b}$, and N.
Sterpi$^{a}$ \\
{\em $^a$ Dipartimento di Fisica \lq A. Volta\rq, Universit\`a degli 
Studi di Pavia} \\
{\em via A. Bassi 6, I-27100 Pavia, Italy}\\
{\em $^b$ Clarendon Laboratory, University of Oxford}\\
{\em Parks Road, OX1 3PU, Oxford, UK}} 
\maketitle
\begin{abstract}\footnotesize
We study both systematic and statistical errors in radiation density 
matrix measurements. First we estimate the minimum number of scanning 
phases needed to reduce systematic errors below a fixed 
threshold.
Then, we calculate the statistical errors, intrinsic in the 
procedure that gives the density matrix. We present a detailed study of 
such errors versus the detectors quantum efficiency $\eta$ and 
the matrix indexes in the number representation,
for different radiation states.
For unit quantum efficiency, and for both 
coherent and squeezed states, the statistical errors 
of the diagonal matrix elements saturate for large $n$. 
On the contrary, off-diagonal errors increase with the distance from the 
diagonal.
For non unit quantum efficiency the statistical errors along the diagonal 
do not saturate, and increase dramatically versus both
$1-\eta$ and the matrix indexes. 
\end{abstract} 

\section{Introduction}

The possibility of investigating quantum radiation states by homodyne
detection techniques recently raised much interest\cite{Rassegnal}. 
In particular, progress has been made on the determination of an 
exact method to detect the density matrix directly from homodyne 
measurements, in any representation, 
without resorting to any smoothing or filtering procedure of experimental 
data\cite{Dariano94,Dariano95,Leonhardt95}. Such method can be basically 
summarized as follows. By means of homodyne detection, the field quadrature 
$\hat x_\phi=(a^{\dag}e^{i\phi}+ae^{-i\phi})/2$ is measured at any desired 
phase shift $\phi$ with respect to the local oscillator ($a^{\dag}$ and $a$ 
are the creation and annihilation operators of the field mode). 
Then, the density matrix elements are obtained by averaging functions,
called ``kernel functions'' (or ``pattern functions''), on experimental  
data. We call this procedure ``homodyning the density matrix''
\cite{Rassegnad}, to  
distinguish it from the previously used methods, where the density matrix was 
reconstructed after evaluating the Wigner function as an intermediate step
(the celebrated ``quantum tomography'' \cite{Vogel,Raymer,kuhn}).
The present method takes into account the detectors quantum efficiency,
which must be greater than 0.5 for measuring the density matrix in the 
number representation\cite{Dariano95}. 

In this paper, we numerically evaluate the measurement accuracy and
the statistical errors in homodyning the density matrix, for both unit 
and non unit detectors quantum efficiency $\eta$. In Section \ref{s:method} 
we briefly recall the direct method of homodyning the 
density matrix. Since each matrix element is given by an 
integral over scanning phases, whose number is necessarily finite, non 
negligible systematic errors arise if the number of phases 
is not large enough. Thus, in Section \ref{s:systematic} 
we numerically estimate a lower value $f_0$ for the number of phases 
$f$, needed 
for an accurate measurement of a state, and we study the dependence of $f_0$
on the kind of state. We also investigate the convergence of the reconstructed 
matrix elements towards 
their respective theoretical values as functions of $f$. 
In Section \ref{s:statistical} we introduce the statistical errors
of the measured matrix elements. 
We study the errors as functions of the matrix indexes and of the quantum 
efficiency, for both coherent and squeezed states. We show that, for
$\eta=1$, the statistical errors of the diagonal matrix elements
saturate for large $n$. This result is also analytically obtained 
after introducing an asymptotic approximation for the kernel functions. 
The off-diagonal errors increase with the distance from the 
diagonal. For $0.5<\eta<1$, we show that the statistical errors along 
the diagonal do not saturate, and increase dramatically versus both
$1-\eta$ and the matrix indexes. Due to such statistical errors, it is
not convenient to use the measured density matrix elements to evaluate 
the expectation values of generic observables. Therefore, in the end of 
Section \ref{s:statistical} we consider the possibility of {\em homodyning 
the observable}, i.e. measuring directly the expectation value of an 
observable by experimentally sampling an appropriate kernel function. 
In particular, we consider the number of photons, and we calculate the 
precision of this kind of measurement. 
In Section \ref{s:conclusions} we conclude 
the paper, and in Appendix we report some useful calculations in detail.
 
\section{Homodyning the density matrix}\label{s:method}

We briefly recall the method for homodyning the radiation density matrix 
$\hat \rho$. Our starting point is the operator identity\cite{Dariano95}
\begin{eqnarray}
\hat \rho = 
\int_0^{\pi} \frac{\mbox{d}\phi}{\pi}
\int_{-\infty}^{+\infty} \mbox{d}k \, \frac{|k|}{4} \,
\mbox{Tr}[\hat \rho e^{ik\hat x_\phi}]\,e^{-ik\hat x_\phi} \, ,
\label{identity}
\end{eqnarray}
where the field quadrature is defined as $\hat x_\phi=(a^{\dag}e^{i\phi}
+ae^{-i\phi})/2$ and $\phi$ is the phase with respect to the local 
oscillator.
The trace in Eq.(\ref{identity}) can be written in terms of quadrature
probability distributions $p_\eta(x,\phi)$ at phase $\phi$: for 
detectors quantum efficiency $\eta<1$, such distributions are related to the 
ideal one ($\eta=1$) by a Gaussian convolution so that in terms of 
characteristic functions one has
\begin{eqnarray}
\mbox{Tr}[\hat \rho e^{ik\hat x_\phi}]=
e^{\frac{1-\eta}{8\eta}k^2}
\int_{-\infty}^{+\infty} 
\mbox{d}x \, p_\eta(x,\phi) \, e^{ikx}\, .
\end{eqnarray}
After exchanging integrals over $k$ and $x$, Eq.(\ref{identity}) reads  
\begin{eqnarray}
\hat \rho = 
\int_0^{\pi} \frac{\mbox{d}\phi}{\pi} \, 
\int_{-\infty}^{+\infty} 
\mbox{d}x \, p_\eta(x,\phi) \, 
\hat K^{(\eta)}_{\phi}(x) \, .
\label{rho}
\end{eqnarray}
In Equation (\ref{rho}), the kernel operator $\hat K^{(\eta)}_{\phi}(x)$ is
\begin{eqnarray}
\hat K^{(\eta)}_{\phi}(x)= 
e^{i\phi a^{\dag}a}\hat\nu^{(\eta)}(x)
e^{-i\phi a^{\dag}a}
\label{kernelop} 
\end{eqnarray}
with  
\begin{eqnarray}
\hat \nu^{(\eta)}(x)= 
\int_{-\infty}^{+\infty} 
\mbox{d}k \frac{|k|}{4}\, e^{-\frac{2\eta-1}{8\eta}k^2+ikx}
\, :e^{-ik\frac{a^{\dag}+a}{2}}: 
\label{nuint}
\end{eqnarray}
(where $:\::$ denotes normal ordering). The operator $\hat \nu^{(\eta)}(x)$ 
can also be written as 
\begin{eqnarray}
\hat \nu^{(\eta)}(x)=
\partial_x\hat \mu^{(\eta)}(x) \, ,
\label{nu} 
\end{eqnarray}
with
\begin{eqnarray}
\hat \mu^{(\eta)}(x)=
\sqrt{2}\chi
:e^{-\frac{a^{\dag}+a}{2}\partial_x}:
e^{-2\chi^2 x^2} 
\int_0^{\sqrt{2}\chi x} \mbox{d}t \, e^{t^2}
\label{mu} 
\end{eqnarray}
and $\chi=\sqrt{\eta/(2\eta-1)}$. Notice that, equivalently, one has
\begin{eqnarray}
\hat \mu^{(\eta)}(x)=
\sqrt{2}\chi
e^{-\frac{a^{\dag}+a}{2}\partial_x} e^{-\frac{1}{8}\partial^2_x} 
e^{-2\chi^2 x^2}
\int_0^{\sqrt{2}\chi x} \mbox{d}t \, e^{t^2} \, , 
\end{eqnarray}
where the antidiffusion operator $\exp(-\partial^2_x/8)$ is due to
normal ordering in Eq.(\ref{mu}). 

The matrix elements 
$\rho(n,m)=\langle n|\hat\rho|m\rangle$ are evaluated by averaging 
the kernel functions [i.e. the matrix elements of the kernel operator
$\hat K^{(\eta)}_\phi(x)$] calculated for random homodyne outcomes.
As the experimental data are distributed according to the 
probability $p_\eta(x,\phi)$, such average gives a measurement of the density
matrix. In other words, the density matrix is measured by experimentally 
sampling the kernel functions. 

The kernel functions for homodyning the density matrix are reported in the 
following. We have carried out our analysis in the 
number representation, for $\eta$ greater than the 
lower bound 0.5 (it has been shown that $\eta=0.5$ is a
universal lower bound for any representation\cite{Rassegnad}). 

\subsection{Unit quantum efficiency}

For $\eta=1$, Eq. (\ref{mu}) reads:
\begin{eqnarray}
\hat \mu(x)=
\sqrt{2} 
:e^{-\frac{a^{\dag}+a}{2}\partial_x}:
e^{-2x^2} \int_0^{\sqrt{2}x} \mbox{d}t \, e^{t^2}
\label{mu1}
\end{eqnarray}
A simple and fast algorithm can be derived after writing the matrix
elements  
$\langle n|\hat\mu(x)|m\rangle$ in factorized form. This technique was
first introduced by Richter\cite{Richter} for diagonal matrix elements,
and was later generalized to off-diagonal matrix elements by Leonhardt 
{\em et al.}\cite{Leonhardt}. In Appendix we present a simple and 
alternative derivation which, in our opinion, is useful for further 
developments. The kernel functions are calculated from Eq.(\ref{fact})
and they read 
\begin{eqnarray}
\langle m+d|\hat K^{(\eta)}_{\phi}(x)|m\rangle
&=& e^{id\phi} \left[ 4x \, u_m(x)v_{m+d}(x) 
- 2\sqrt{m+1}\,u_{m+1}(x)v_{m+d}(x) \right.
\nonumber \\
&-& \left. 2\sqrt{m+d+1}\,u_{m}(x)v_{m+d+1}(x) \right] \, ,
\end{eqnarray}
where the functions $u_j(x)$ and $v_{j}(x)$ are              
respectively the normalizable and the non normalizable eigenfunctions 
of the harmonic oscillator (corresponding to eigenvalue $j$).

\subsection{Non unit quantum efficiency}

For $\eta<1$, no factorization algorithm is known at present. In this
case, from Eq.(\ref{nuint}) we obtain the following form for the kernel 
functions\cite{Dariano95}:
\begin{eqnarray} 
&&\langle m+d|\hat K^{(\eta)}_\phi(x)|m\rangle
= e^{id\phi} 2\chi^{d+2}\sqrt{\frac{m!}{(m+d)!}}e^{-\chi^2x^2} 
\\ \nonumber 
&&\times \sum_{\nu=0}^{m}\frac{(-)^\nu}{\nu!}
\left(\begin{array}{c} m+d \\ m-\nu \end{array}\right)(2\nu+d+1)!
\chi^{2\nu}\mbox{Re}\left\{(-i)^d D_{-(2\nu+d+2)}(-2i\chi x)\right\} \, ,
\label{kernel}
\end{eqnarray}
where $D_j(\sigma)$ denotes the parabolic cylinder function. 

\section{Systematic errors}\label{s:systematic}

In Equation (\ref{rho}) the density matrix is given by an integral over 
the phase $\phi$ with respect to the local oscillator. In order to avoid 
any systematic error, one should homodyne the density matrix at perfectly 
random phases. This is the case of the experimental method of Munroe 
{\em et al.}\cite{Munroe}, where the photon number probability distribution 
is measured by homodyne detection: 
in such measurement no knowledge of the phase is needed, because the 
diagonal kernel functions are independent of $\phi$. However, for measuring 
off-diagonal matrix elements the knowledge of the random phase is essential, 
and it is difficult to achieve. In such situation, the phase integral
is usually performed by a phase scanning, as in Ref.\cite{Raymer}. 
An insufficient number 
of phases generates systematic errors, leading to values for the 
density matrix elements that are far from the true values. Therefore, 
in the experimental determination of the density matrix one has to 
eliminate these systematic errors as the first step.

The criterion adopted here to establish the degree of accuracy in a 
measurement is based on the {\em absolute} deviation of the measured matrix 
elements from the ``true'' matrix elements. For each $\rho(n,m)$, obtained 
from Eq.(\ref{rho}), we calculate the absolute deviation 
\begin{eqnarray}
\epsilon(n,m)=|\rho(n,m)-\rho_t(n,m)| \, , 
\label{epsilon}
\end{eqnarray}
where $\rho_t(n,m)$ 
is the true (theoretical) density matrix.
For fixed state, the set $\{\epsilon(n,m)\}$ ($n,m=0,1,...$) 
depends on the number of scanning phases $f$ used in the experiment
(the number of experimental data per scanning phase is kept fixed).
We have an accurate matrix measurement when the 
maximum deviation is reduced below a fixed threshold, for example 
\begin{eqnarray}
\epsilon=\max_{n,m} \{\epsilon(n,m)\} < 10^{-4}
\quad \quad \quad (n,m=0,1...) \, .
\label{cond}
\end{eqnarray}
Now, let us show how the accuracy depends on $f$ for different radiation 
states. 

The measurement accuracy increases with $f$. We expect that the more a 
radiation state is either displaced or ``asymmetrically'' distributed in 
phase space, the higher the number $f$ must be. This is indeed the case. In 
Fig.\ref{f:nf} we show the minimum number of phases $f_0$ needed for an 
accurate measurement of coherent and squeezed states. This number increases 
with both the average number of photons $\langle \hat n \rangle$ and the 
squeezing parameter $r$\cite{notesqpar}. 

We point out that far off-diagonal kernel functions
oscillate very fast as functions of $\phi$, thus the larger the matrix
dimension, the larger $f_0$. However, the main result,  
i.e. the increase of $f_0$ with $\langle \hat n \rangle$ and $r$, does not 
change. Indeed, both an increase and a decrease of the matrix dimension 
merely shift the plot in Fig.\ref{f:nf} towards either higher or lower 
values of $f_0$. In the following we set $n_{max}=47$.  

A comment about our choice for the accuracy criterion is now in order. 
Our purpose is to show the dependence of $f_0$ on the average energy and 
on the ``asymmetry'' in the phase space. This is achieved by calculating 
the absolute deviations $\{\epsilon(n,m)\}$: indeed, the systematic 
errors are independent of the size of the theoretical matrix element.

Finally, we briefly examine the dependence on the number of phases $f$ 
for measurements of individual matrix elements.

We expect that for off-diagonal matrix elements the number of phases needed 
for an accurate measurement is larger than for diagonal ones, due to 
faster oscillations of the integrand in Eq.(\ref{rho}) versus
$\phi$. For coherent states this is 
generally true, as shown for example in Fig.\ref{f:devco}, where 
$\epsilon(5,5)< 10^{-4}$ for $f\geq$14, and $\epsilon(18,5)< 10^{-4}$ 
for $f\geq$24. For squeezed states the behavior on the distance from the 
diagonal is more complicated. In many cases 
the same result of coherent states is found, see for example 
Fig.\ref{f:devsq_5}, where the diagonal element $\rho(5,5)$ converges faster
than $\rho(10,5)$ and $\rho(15,5)$ for large enough $f$. However, there 
are exceptions to this behavior. As an example, in Fig.\ref{f:devsq_10} 
we show the asymptotically slower convergence of $\rho(10,10)$ with respect 
to $\rho(10,9)$ and $\rho(10,0)$.

\section{Statistical errors}\label{s:statistical}

The statistical errors on the measured matrix elements are calculated 
in terms of the errors on real and imaginary parts of the matrix. For a 
matrix element $\rho(n,m)$ the real part of the statistical variance
is defined as
\begin{eqnarray}
\mbox{Re}^2\{\sigma(n,m)\}
&=&\int_0^{\pi} \frac{\mbox{d}\phi}{\pi} \, 
\int_{-\infty}^{+\infty} 
\mbox{d}x \, p_\eta(x,\phi) \, 
\mbox{Re}^2\{\langle n|\hat K^{(\eta)}_\phi(x)|m\rangle\} 
\nonumber \\
&-&\mbox{Re}^2\{\rho(n,m)\} 
\label{sigma}
\end{eqnarray}
and analogously for the imaginary part. The 
experimental error of the measurement is obtained by rescaling the
amplitudes $|\sigma(n,m)|$ by a factor $1/\sqrt{N}$, where $N$ is the 
total number of experimental data. For simplicity, hereafter
the quantity $\sigma(n,m)$ will be called statistical error.
The statistical errors turn out to be independent of $f$ if $f>f_0$.
Thus, we focus attention
on the general features of the set $\{\sigma(n,m)\}$ for 
different radiation states, at fixed $f$. First we show the results for unit 
quantum efficiency $\eta$, later we will consider the dependence on $\eta$.

\subsection{General features for unit quantum efficiency} 

For coherent and squeezed radiation states, the real and imaginary parts 
of the statistical errors exhibit a similar behavior as functions of the 
matrix indexes (with the major exception of the matrix diagonal, where 
obviously $\mbox{Im}\{\sigma(n,n)\}\equiv 0$).
Thus, without loss of generality, we can show our results in terms of 
the amplitudes $|\sigma(n,m)|$. 

In Fig.\ref{f:co} we report the matrix of errors $|\sigma(n,m)|$ 
for a coherent state with $\langle \hat n \rangle=4$.   
The contour plot shows that errors increase with the distance $d=n-m$ from
the diagonal. This is related to the analytical form of the kernel operator. 
In particular, for fixed $\phi$, all the kernel functions are oscillating 
functions of $x$\cite{QO}. 
Moreover, for increasing $d$ the oscillations become 
faster and the oscillation range slowly increases. If a kernel function 
oscillates fast, its statistical average becomes more
sensitive to fluctuations of experimental
data and, therefore, the statistical errors must increase versus $d$. 

The contour plot also emphasizes the ``saddle region'' around the diagonal, 
suggesting that the statistical errors for measured diagonal matrix elements 
saturate to a value independent of $n$ for large enough $n$. 
This is shown more clearly in Fig.\ref{f:codiag}. 
Such a remarkable feature is general. In fact, it is 
independent on the energy $\langle \hat n \rangle$ and, more important, 
it holds for any state. Noticeably, the 
limiting value $|\sigma(n,n)|=\sqrt{2}$ does not depend the degree of
squeezing. 
The reason for such saturation is due to the analytic
form of the diagonal kernel functions. 
Indeed, the larger $n$ is, the faster the kernel functions oscillate vs. 
$x$ and the errors must increase with $n$. On the other hand, for
$d=0$ the range of oscillation is fixed between $-2$ and 2, 
thus the diagonal errors are 
bounded, and hence they must saturate. These
considerations are confirmed by considering the explicit form of the 
statistical errors, as given by Eq.(\ref{sigma}). In particular, 
from Eq.(\ref{sigma}) we can extract the relevant contribution for large
$n$ upon considering that the kernel functions
oscillate fast in the region where $p(x,\phi)$ is sizeable.
Moreover, $p(x,\phi)$ has a Gaussian decay, whereas the
kernel functions decrease as a power of $x$. Thus,  
\begin{eqnarray}
|\sigma(n,n)|\simeq
\left\{\int_0^{\pi} \frac{\mbox{d}\phi}{\pi} \, 
\int_{-\infty}^{+\infty} 
\mbox{d}x \, p(x,\phi) \, 4\cos^2(k_n x) \right\}^{1/2}\,.
\label{appr}
\end{eqnarray}
For large values of $n$, $k_n\to\infty$: 
if $p(x,\phi)$ can be considered constant over a cycle $\Delta
x={\pi}/{k_n}$, 
the integral over $x$ in Eq. (\ref{appr}) gives just the average of
$\cos^2(k_n x)$, which leads to 
\begin{eqnarray}
|\sigma(n,n)|\simeq\sqrt{2} \, .
\end{eqnarray}
If very squeezed states are considered [i.e. with very sharp
$p(x,\phi)$] the errors will saturate for larger $n$.
In Fig.\ref{f:sq} we show $|\sigma(n,m)|$ for a squeezed state with 
$\langle \hat n \rangle=4$ and $r=1$: the plot is quite different
from Fig.\ref{f:co}, but the diagonal errors still saturate. 

\subsection{Dependence on the quantum efficiency}

The influence of the quantum efficiency $\eta$ on $|\sigma(n,m)|$ 
is very strong. Indeed, for non unit quantum efficiency of detectors 
the behavior 
of the kernel functions (\ref{kernel}) changes dramatically: for fixed 
$n$ and $m$, the oscillation range increases very rapidly as $\eta$ 
approaches the lower bound $\eta=0.5$, and the resulting errors 
increase rapidly as well. The growth rate is different for 
different matrix elements: as an example, in Fig.\ref{f:diag_eta} we show 
some diagonal errors as functions of quantum efficiency 
(for a coherent state). Furthermore, the diagonal errors no longer 
saturate for large values of $n$. Very similar results are found 
for squeezed states. In particular, the growth rate of diagonal errors 
$|\sigma(n,n)|$ vs. $1-\eta$ is slightly larger than for coherent states.  
The diagonal errors for a squeezed state are shown in 
Fig.\ref{f:sqdiag_eta} for different values of $\eta$.

For fixed $\eta<1$, the oscillation range of the kernel functions increases 
with both $n$ and the distance $d$ from the diagonal. Thus, for increasing 
$n$ and $d$ the statistical errors increase. For example, we consider 
$\eta=0.99$: after a comparison between Fig.\ref{f:co99} 
and Fig.\ref{f:co}, one can see that the open contour levels for 
$|\sigma(n,m)|$ close, and any errors saturation disappears.
Figure \ref{f:co99} shows that drastic modifications arise with respect 
to the ideal case for $\eta=1$. This means that, already for $\eta=0.99$, 
in order to have the same experimental errors on the
measurement of the density matrix, the number of data must be much
larger than in the ideal case. 

\subsection{Precision of homodyning observables}

From the measured density matrix, one can evaluate the probability 
distributions of operators that are functions of the field operators 
$a$ and $a^\dagger$. Thus, by means of homodyne experimental data it is 
possible to obtain indirect measurements of observables. However, 
for some observables the propagation law of statistical errors leads 
to additional noise with respect to direct detection. In some
cases, such indirect detection through the density matrix can be overcome 
by a more convenient procedure, namely {\em homodyning the observable}. 
By {\em homodyning the observable} we mean the experimental sampling of an
appropriate kernel function, which directly gives the expectation value of the
desired observable. 

We consider, as an example, the homodyne measurement 
of the mean photon number $\langle \hat n \rangle$. From Eq.(\ref{identity}), 
$\langle \hat n \rangle$ is expressed as
\begin{eqnarray}
\langle \hat n \rangle = 
\int_0^{\pi} \frac{\mbox{d}\phi}{\pi}
\int_{-\infty}^{+\infty} \mbox{d}x \, p(x,\phi) \, F(x,\phi)
\end{eqnarray}
where 
\begin{eqnarray}
F(x,\phi) \equiv F(x)=\sum_{n=0}^\infty n 
\int_{-\infty}^{+\infty} \mbox{d}k \, 
\frac{|k|}{4} \, e^{-\frac{1}{8}k^2+ikx} \, 
L^{(0)}_n \left( \frac{k^2}{4} \right) 
= 2x^2-\frac{1}{2} \, .
\label{kernelnum}
\end{eqnarray}
In Equation (\ref{kernelnum}), $L^{(0)}_n$ denote zero-order Laguerre 
polynomials and unit detectors efficiency has been considered.
The statistical fluctuations of the measured mean photon number are given by
\begin{eqnarray}
\sigma^2_{\langle \hat n \rangle}= 
\int_0^{\pi} \frac{\mbox{d}\phi}{\pi}
\int_{-\infty}^{+\infty} \mbox{d}x \, p(x,\phi) \, F^2(x,\phi)
-\langle \hat n \rangle^2 \, . 
\label{errn}
\end{eqnarray}
and $\sigma_{\langle \hat n \rangle}$ is the statistical 
error for homodyning the mean photon number. 
The precision $\epsilon_{\langle\hat n\rangle}$ of this
homodyne measurement is defined by the relation
\begin{eqnarray}
\epsilon^2_{\langle\hat n\rangle}=
\sigma^2_{\langle \hat n \rangle}-\langle\Delta\hat n^2\rangle \, ,
\end{eqnarray}
where $\langle\Delta\hat n^2\rangle$ is the intrinsic
quantum uncertainty
\begin{eqnarray}
\langle\Delta\hat n^2\rangle 
&\equiv& \langle \hat n^2 \rangle - \langle\hat n\rangle^2
= \langle a^{\dag 2} a^2\rangle + \langle \hat n \rangle 
- \langle \hat n \rangle^2 \, .
\label{dn}
\end{eqnarray}
The uncertainty $\langle\Delta\hat n^2\rangle$ can be expressed in terms 
of quadrature probability distributions: after calculating the kernel 
function for operator $ a^{\dag 2} a^2$\cite{RichterPRA}, 
Eq.(\ref{dn}) reads
\begin{eqnarray}
\langle\Delta\hat n^2\rangle = \int_0^{\pi} \!\frac{\mbox{d}\phi}{\pi}
\int_{-\infty}^{+\infty} \!\!\mbox{d}x \, p(x,\phi) \, 
\left\{\frac{8}{3}x^4-2x^2\right\} \, -\langle \hat n \rangle^2 \, .
\end{eqnarray}
In conclusion, the precision for homodyning the photon number is
\begin{eqnarray}
\epsilon_{\langle\hat n\rangle}=
\frac{1}{\sqrt{2}} \left( \langle\Delta\hat n^2\rangle + 
\langle\hat n\rangle^2 + \langle\hat n\rangle +1 \right)^{1/2} \, .
\end{eqnarray}

\section{Conclusions}\label{s:conclusions}

We analyzed both systematic and statistical errors for homodyne detection   
of the density matrix of light. Such detection is performed by
suitably processing  homodyne experimental data. We studied 
the behavior of systematic errors as functions of the number of
scanning phases $f$. We calculated the lower bound
for $f$, needed for an accurate matrix measurement. We found that 
this lower bound increases with both the mean photon number and the 
``asymmetry'' in phase space of the state. Then we considered the
statistical errors corresponding to the data average that gives 
each matrix element. Noticeably, for unit quantum efficiency of the
detectors the diagonal errors $\sigma(n,n)$ of the matrix elements 
$\rho(n,n)$ ``saturate'' to the fixed value $\sqrt{2}$ for large enough $n$. 
This feature is independent of the degree of squeezing. The 
off-diagonal errors increase with the distance from the diagonal. 
If detectors quantum efficiency is decreased, the errors increase for
each matrix element. In particular, any saturation effects disappear. 
Finally, we considered the homodyne detection of
the mean photon number, that is achieved by sampling 
an appropriate kernel function, and we analytically evaluated the 
precision of such measurement. We think that the results 
presented here are relevant from a fundamental
point of view and provide the experimentalist with important 
information on the behavior of errors in homodyning the density matrix.

\section{Appendix}\label{s:Appendix}

The factorization of the matrix element $\langle m+d|\hat\mu(x)|m\rangle$
is performed in two steps. By setting $n=m+d$ we obtain 
\begin{eqnarray}
\langle m+d|\hat\mu(x)|m\rangle
&=& \sqrt{\frac{m!}{(m+d)!}}
\sum_{\nu=0}^m 
\left(\begin{array}{c} m+d \\ \nu+d\end{array}\right)
\frac{1}{\nu!} 
\nonumber \\
&\times& \left(- \frac{\partial_x}{2}\right)^{2\nu+d}
\sqrt{2} e^{-2x^2} \int_0^{\sqrt{2}x} \mbox{d}t \, e^{t^2} \, .
\label{afact}
\end{eqnarray}
Then, the derivatives with respect to $x$ and the summation are evaluated
as follows. We introduce the ``seed functions''  
\begin{eqnarray}
u_0(x)&=&\left(\frac{2}{\pi}\right)^{1/4} e^{-x^2} \\
v_0(x)&=&\left(2\pi\right)^{1/4} e^{-x^2}
\int_0^{\sqrt{2}x} \mbox{d}t \, e^{t^2} 
\end{eqnarray}
that generate two sets of functions $\{u_j(x)\}$ and $\{v_j(x)\}$ for
$j=0,1,2...$, as
\begin{eqnarray}
u_j(x)=\frac{1}{\sqrt{j!}}
\left( x-\frac{\partial_x}{2}\right)^j u_0(x) \\ 
v_j(x)=\frac{1}{\sqrt{j!}}
\left( x-\frac{\partial_x}{2}\right)^j v_0(x) \, .
\label{def} 
\end{eqnarray}
By means of the following identity between operators
\begin{eqnarray}
\partial_x u_0(x) = u_0(x)(\partial_x-2x)\, , 
\end{eqnarray}
we obtain 
\begin{eqnarray}
\left(-\frac{\partial_x}{2}\right)^d u_0(x)v_0(x)=
\sqrt{d!} \, u_0(x)v_d(x) \, .
\label{der0} 
\end{eqnarray}
As noticed in Ref. \cite{Leonhardt},
the functions $\{u_j(x)\}$ and $\{v_j(x)\}$ are 
respectively the normalizable and the non normalizable eigenfunctions 
of the harmonic oscillator (corresponding to eigenvalue $j$). Thus, by using 
the standard recursion relations for the harmonic oscillator eigenfunctions, 
we can easily demonstrate the following identity\cite{Note1}:
\begin{eqnarray}
\frac{1}{\nu!}\left(- \frac{\partial_x}{2}\right)^{2\nu}
u_0(x)v_d(x)=
\sum_{j=0}^\nu 
\sqrt{\left(\begin{array}{c} j+d \\ j \end{array}\right)} 
(-1)^{\nu-j} 
\left(\begin{array}{c} \nu+d \\ j+d\end{array}\right)
u_j(x)v_{j+d}(x) \, .
\label{der}
\end{eqnarray}
After substituting (\ref{der0}) and (\ref{der}) in Eq.(\ref{afact}), we 
obtain the factorized formula 
\begin{eqnarray}
\langle m+d|\hat\mu(x)|m\rangle=\langle m|\hat\mu(x)|m+d\rangle=
u_m(x)v_{m+d}(x) \, ,
\label{fact}
\end{eqnarray}
where we use the fact that $\hat\mu(x)$ is real selfadjoint. 
Finally, the kernel functions are obtained using Eqs. 
(\ref{kernelop}),(\ref{nu}) and (\ref{fact}).

\newpage 

\begin{figure}
\caption{Minimum number of scanning phases $f_0$ required by the condition 
$\epsilon<10^{-4}$ vs. the mean number of photons $\langle \hat n \rangle$ 
for coherent states (circles), squeezed states with $r=0.6$ (triangles), 
and $r=1$ (squares). [The matrix dimensions are fixed to $n_{max}=47$.]}
\label{f:nf}   
\end{figure}
\begin{figure}
\caption{Absolute deviation $\epsilon(n,m)$ vs. $f$ for a coherent state 
with $\langle \hat n \rangle=4$: $(n,m)=(5,5)$ (circles), $(n,m)=(10,5)$ 
(triangles), $(n,m)=(18,5)$ (squares). The theoretical matrix elements are
$\rho_t(5,5)=0.15629$, $\rho_t(10,5)=0.02876$, $\rho_t(18,5)=0.00017$.}
\label{f:devco}   
\end{figure}
\begin{figure}
\caption{Absolute deviation $\epsilon(n,m)$ vs. $f$ 
for a squeezed state with $\langle \hat n \rangle=4$, $r=1$: $(n,m)=(5,5)$ 
(circles), $(n,m)=(10,5)$ (triangles), $(n,m)=(15,5)$ (squares).
The theoretical matrix elements are $\rho_t(5,5)=0.04182$, 
$\rho_t(10,5)=0.03231$, $\rho_t(15,5)=0.01852$.}
\label{f:devsq_5}   
\end{figure}
\begin{figure}
\caption{Absolute deviation $\epsilon(n,m)$ vs. $f$ for a squeezed 
state with $\langle \hat n \rangle=4$, $r=1$: $(n,m)=(10,10)$ 
(circles), $(n,m)=(10,9)$ (triangles), $(n,m)=(10,0)$ (squares).
The theoretical matrix elements are $\rho_t(10,10)=0.02495$, 
$\rho_t(10,9)=0.02418$, $\rho_t(10,0)=0.09307$.}
\label{f:devsq_10}   
\end{figure}
\begin{figure}
\caption{Statistical error amplitudes $|\sigma(n,m)|$ for a coherent state 
with $\langle \hat n \rangle=4$ ($\eta=1$).}
\label{f:co}   
\end{figure}
\begin{figure}
\caption{$|\sigma(n,n)|$ for: coherent state 
with $\langle \hat n \rangle=4$ (circles), squeezed state with 
$\langle \hat n \rangle=4$, $r=1$ (triangles) ($\eta=1$).}
\label{f:codiag}   
\end{figure}
\begin{figure}
\caption{$|\sigma(n,m)|$ for a squeezed state
with $\langle \hat n \rangle=4$, $r=1$ ($\eta=1$).}
\label{f:sq}   
\end{figure}
\begin{figure}
\caption{$|\sigma(n,n)|$ vs. $1-\eta$ 
for $n=0,2,5,15$ on a semilogarithmic scale (for a coherent state with 
$\langle \hat n \rangle=4$). The quantum efficiencies are 
$\eta=1,0.99,0.97,0.95,0.9$.}
\label{f:diag_eta}   
\end{figure}
\begin{figure}
\caption{$|\sigma(n,n)|$ for a squeezed state 
with $\langle \hat n \rangle=4$, $r=1$ for: $\eta=1$ (circles), $\eta=0.99$ 
(triangles), $\eta=0.97$ (squares), $\eta=0.95$ (rhombi), $\eta=0.9$ 
(stars).}
\label{f:sqdiag_eta}
\end{figure}
\begin{figure}
\caption{$|\sigma(n,m)|$ for a coherent 
state with $\langle \hat n \rangle=4$ for quantum efficiency 
$\eta=0.99$.}
\label{f:co99}   
\end{figure}

\newpage 

\begin{thebibliography}{99}
\bibitem[*]{dar}E-mail: dariano@pv.infn.it
\bibitem{Rassegnal} A review on methods to measure quantum states 
of radiation is given by U. Leonhardt and H. Paul, Prog. Quantum Electron.
{\bf 19}, 89 (1995).
\bibitem{Dariano94} G.M. D'Ariano, C. Macchiavello, and M.G.A. Paris,
Phys. Rev. A {\bf 50}, 4298 (1994).
\bibitem{Dariano95} G.M. D'Ariano, U. Leonhardt, and H. Paul, Phys. Rev. A 
{\bf 52}, R1801 (1995).
\bibitem{Leonhardt95} U. Leonhardt, H. Paul, and G.M. D'Ariano, Phys. Rev. A 
{\bf 52}, 4899 (1995).  
\bibitem{Rassegnad} A recent review on both the new direct method for 
reconstructing the density matrix from homodyne data and the previous 
tomographic methods is given by G. M. D'Ariano, {\em Measuring quantum
states}, in {\em Concepts and Advances in Quantum Optics and Spectroscopy 
of Solids}, ed. by T. Hakioglu and A.S. Shumovsky (Kluwer, Amsterdam, 
in press).
\bibitem{Vogel} K. Vogel and H. Risken, Phys. Rev. A {\bf 40}, 2847 (1989).
\bibitem{Raymer} D.T. Smithey, M. Beck, M.G. Raymer, and A. Faridani, 
Phys. Rev. Lett. {\bf 70}, 1244 (1993).
\bibitem{kuhn} H. K\"{u}hn, D.-G. Welsch, W. Vogel, J. Mod. Opt. {\bf
41}, 1607 (1994).
\bibitem{Richter} Th. Richter, Phys. Lett. A {\bf 211}, 327 (1996). 
\bibitem{Leonhardt} U. Leonhardt, M. Munroe, T. Kiss, Th. Richter, 
and M. G. Raymer, Opt. Comm. {\bf 127}, 144 (1996). 
\bibitem{Note1} Equation \protect{(\ref{der})} is demonstrated by 
means of the recursion relation 
\protect{\begin{eqnarray}
&&\frac{\partial^2_x}{4}u_m(x)v_{m+d}(x)=
\sqrt{m(m+d)}u_{m-1}(x)v_{m-1+d}(x)\nonumber\\
&&-\left(1+2m+d\right)u_m(x)v_{m+d}(x)
+\sqrt{(m+1)(m+1+d)}u_{m+1}(x)v_{m+1+d}(x)\, .\nonumber
\end{eqnarray}}
\bibitem{notesqpar} For squeezing parameter $r$, one has 
\protect{$\langle \hat n \rangle - \sinh^2r =|\langle a \rangle|^2$}. 
\bibitem{Munroe} M. Munroe, D. Boggavarapu, M. E. Anderson, and 
M. G. Raymer, Phys. Rev. A {\bf 52}, R924 (1995). 
\bibitem{QO} G.M. D'Ariano, Quantum Semiclass. Opt. {\bf 7}, 693 (1995).
\bibitem{RichterPRA} Th. Richter, Phys. Rev. A {\bf 53}, 1197 (1996). 
\end{thebibliography}
\end{document}